\documentclass[12pt]{article} 
\input epsf.tex 
\def\be{\begin{equation}} \def\ee{\end{equation}}
\def\bi{\begin{itemize}} \def\ei{\end{itemize}}
\def\bea{\begin{eqnarray}} \def\eea{\end{eqnarray}} \def\ba{\begin{array}}
\def\ea{\end{array}} \def\ben{\begin{enumerate}} \def\een{\end{enumerate}}

\newcommand{\eqn}[1]{(\ref{#1})}

\newcommand{\prl}[3]{Phys. Rev. Lett. {\bf#1} ({#2}) {#3}}

\newcommand{\hepth}[1]{{\tt arXiv:{#1}[hep-th]}}
\newcommand{\arxiv}[1]{{\tt arXiv:{#1}[hep-th]}}

\def\ep{\epsilon}\def\bep{\bar\epsilon}

\def\t{\tau}

\def\br{\nonumber\\}
 \def\ud{\underline}

\begin{document}
{}~
\hfill \vbox{
\hbox{arXiv:2101.nnnn} 
\hbox{\today}}
\break

\vskip 3.5cm
\centerline{\large \bf
Holography and quantum information exchange between systems}

\vskip 1cm

\vspace*{1cm}

\centerline{\sc  Harvendra Singh }

\vspace*{.5cm}
\centerline{ \it  Theory Division, Saha Institute of Nuclear Physics} 
\centerline{ \it  HBNI, 1/AF Bidhannagar, Kolkata 700064, India}
\vspace*{.25cm}

\vspace*{.5cm}

\vskip.5cm
\centerline{E-mail: h.singh [AT] saha.ac.in }

\vskip1cm

\centerline{\bf Abstract} \bigskip

We  estimate the net 
  information exchange 
between  adjacent quantum subsystems holographically
 living on the boundary of 
$AdS$ spacetime. The information exchange
is a real time phenomenon and only after long time interval
it may get saturated. 
Normally we prepare systems
for small time intervals and measure the information exchange
over  finite interval only. 
We find that the information flow between entangled subsystems 
gets reduced if  systems are in excited state whereas
the ground state allows maximum information flow at any given time. 
Especially
for $CFT_2$ we exactly show that a rise in the entropy is 
detrimental to the information exchange by a quantum dot and vice-versa.
We next observe that there is 
a reduction in circuit (CV) complexity too 
 in the presence of excitations for small times.  

\vfill 
\eject

\baselineskip=16.2pt


\section{Introduction}

Since the advent of AdS/CFT \cite{malda} the holography has 
produced simpler answers to many difficult
questions in strongly coupled quantum field theories. 
We consider the phenomenon of information exchange between 
two  quantum subsystems having  common interface. 
The  information exchange between quantum subsystems is 
a real time phenomenon as they would be entangled. 
In quantum theory the information  contained in  state $|\psi>$,
 cannot be destroyed, cloned or even mutated. In
bi-partite systems the
information can be either found in one part of the Hilbert space 
or in the compliment. Thus the quantum subsystems remain always entangled. 
Further it is generally understood
 that the exchange and sharing of quantum information is guided  
by unitarity and locality; see for details \cite{pati,pati1}. 
The time evolution in quantum theory is guided by the Hamiltonian flow. 
In this aspect 
all time dependent flows of isolated quantum system should essentially be
unitary. Under these claims in the black hole evoparation
process (including the Hawking radiation) it is expected that the curve for the
entanglement entropy 
 must bend after the half Page time is crossed \cite{page}. 
This  certainly can hold true when a pure system  
is divided into two small subsystems. 
But for the mixed states, say CFT duals to AdS-black holes,
it is not straight forward to answer this question. 
However,  in recent models by coupling holographic
CFTs to external matter (radiation) bath, and 
involving nonperturbative techniques such as 
replicas and islands \cite{replica19}, people
try to answer some of these questions.\footnote{ See 
 a detailed review on information paradox along
different paradigms in \cite{raju}
and list of references therein.}

In this work we aim to measure in real time exchange of
 quantum information between two {\it adjacent} subsystems (i.e. 
systems having common interface) for the
 field theories dual to AdS black hole spacetimes. The 
information exchange is driven by entanglement between states of the subsystems.
As the quantum systems continuously exchange information  almost
infinite amount of information maybe exchanged over long time intervals. 
This leads to 
an overall information growth in time. The information
exchange grows with time as $\propto({1 \over \delta^{d-2}}- {1 \over t^{d-2}})$ 
for $CFT_d$ ground state. We show in this work that  
the same is true even for  excited state of CFT, but the information
 exchange growth is definitely lowered. Especially for small time 
intervals the loss in 
the information exchange goes quadratically with time. 
In later part of the work we also study  related 
 phenomenon of circuit complexity of quantum systems. The quantum 
complexity is understood to be
a measure of difficulty level in obtaining a target state $|\psi_f>$ 
from  given initial state $|\psi_i>$, by using  minimum number of possible
gates (unitary operations). Here we only explore the complexity 
as the volume (CV) conjecture \cite{susski}. 
 There is an equivalent (CA) conjecture where complexity 
is equated with the SUGRA action evaluated on the Wheeler-De Wit patch 
\cite{CA2}. The related  first law like relation for complexity 
was  proposed in \cite{law1} recently. 
We wish to show that CV complexity can put a bound on 
 information exchange in quantum systems and two phenomena
may be related 
very intrisically.

We also aim  to  understand the
 question about reduction in overall information
exchange when the system is in  excited state and its relationship
with quantum complexity. We find that relative complexity is reduced
for the excited states, although its reduction grows linearly for small time. 
Our approach will be focussed on the  gravity side. 
There have been number of
studies on the related issues; see   \cite{rela1, hashimoto, carmi,
DiGiulio:2020hlz}.

The paper is organized as follows. In section-2 we calculate the 
information exchange between subsystems by calculating the area
of codimension-2 time-dependent RT surface embedded in AdS-BH geometry.
These extremal surfaces are attached to the interface 
between two subsystems. We work  out  the area by 
perturbative procedure up to second order
for small time interval only. Generally, we determine
 that the rate of 
information exchange (measured at the system interface) 
is reduced for the excited CFTs. In section-3 we present 
numerical results  for some special cases only.
These results show expected behaviour at early time. Only
the CFT ground state has  maximum information exchanged 
for a given time. 
We next calculate quantum complexity perturbatively in section-4.
We  observe that there is an 
overall reduction in the growth of
 quantum complexity too in the presence of excitations.  
These results on information exchange and complexity are compared
and related.
Our observation is that complexity  puts a bound on 
the quantum information exchange. 
We conclude with the summary in section-5. (Note added: Our work may have
some overlap with the paper appeared on the arXiv \cite{Ramesh}. 
However their work uses euclidean time frame.)

\section{Information exchange in system entanglement}

We consider  planar $AdS_{d+1}$ geometry 
having a Schwarzschild 
black hole in the IR,  described  by the metric
\bea\label{bst1}
&&ds^2={L^2\over z^2}\left( -{f dt^2}
+dx_1^2+\cdots+dx_{d-1}^2+{dz^2 \over f}\right) 
\eea
with the function $f(z)$ 
\be
 f=1-{z^d\over z_0^d},  
\ee
 where $z=z_0$ is location of the horizon and the 
 boundary  is at $z=0$. 
  $L$ is the radius of curvature of the AdS space, 
which is taken sufficiently large in string length units (i.e.
$L\gg l_s$) so as to suppress the stringy effects. 
The boundary field theory for  $AdS_{d+1}$ spacetime 
describes  $d$-dimensional CFT$_d$ at finite temperature ${d\over \pi z_0}$
\cite{malda}. 
\begin{figure}[ht]
\centerline{\epsfxsize=3.5in
\epsffile{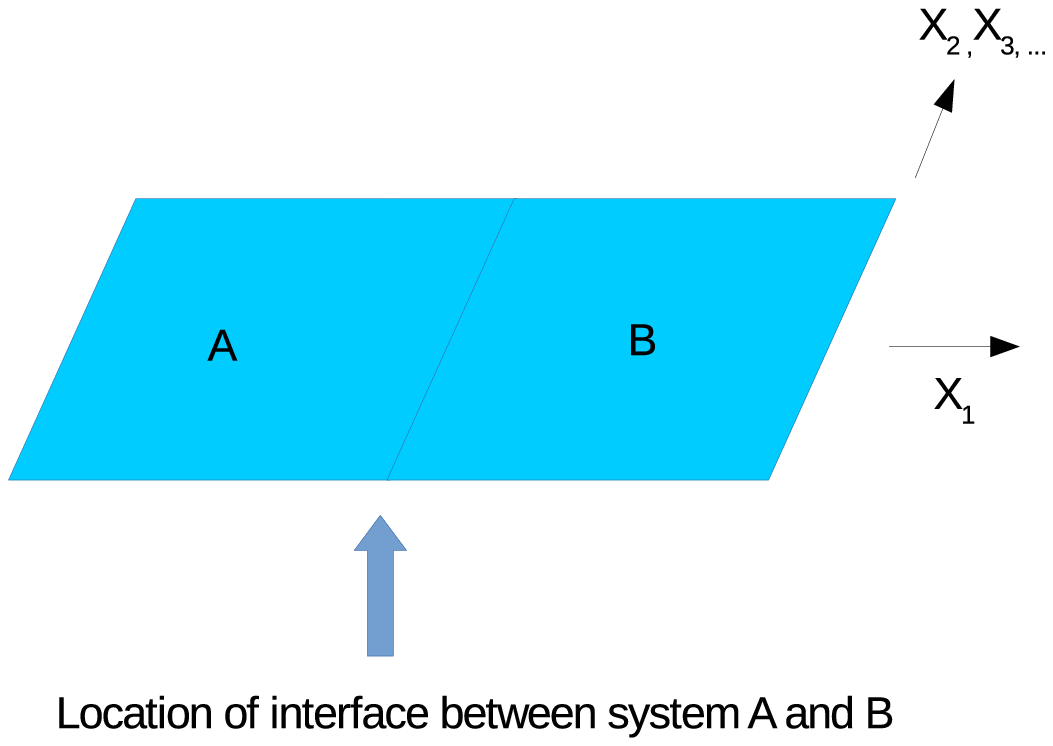} }
\caption{\label{fig1} 
\it  The  interface of two subsystems located at $x_1=x_I$.} 
\end{figure}
\begin{figure}[ht]
\centerline{\epsfxsize=3.5in
\epsffile{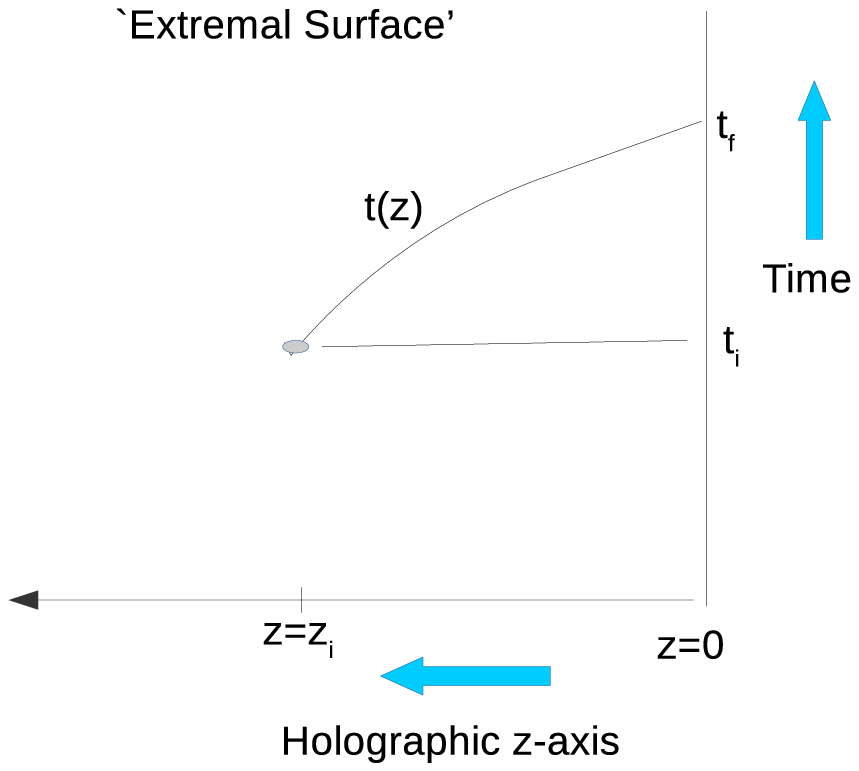} }
\caption{\label{fig2} 
\it The  RT surface between time $t_i$ and $t_f$. Since 
$|\partial_z x^0|\le 1 $ along the extremal curve, 
we get a cusp-point at $z=z_i$. }
\end{figure}
We are  interested in measuring the entanglement information
exchange in the CFT staying
 at the location of interface between two subsystems. 
The interface divided the whole system into two subsystems 
and dynamically is  a $(d-2)$ dimensional spatial plane,
say between system $A$ and the compliment 
$B$, as shown in figure \eqn{fig1}. 
We intend to calculate the area of a  codimension-2 extremal
hyper-surfaces embedded in bulk geometry \eqn{bst1} and anchored precisely 
at the interface, at $x_1=x_I$, between  two large semi-infinite
subsystems.
Since there is homogeneity and  translational invariance, these
results will apply to any constant shift in  
location of the interface ($x_I\to x_I+a$). 
(The subsystems in question can be generalized to a strip.)
 
We now find codimension-2  surface ending at the interface
$x_1=x_I$, and having time-dependent embedding
 as shown in figure \eqn{fig2}. As per holographic 
prescription \cite{RT, HRT}
 generic codimension-2 extremal surfaces can be described by 
an action functional \cite{bousso}
\bea\label{schkl1saa}
 I  &\equiv& 
\int {dz\over z^{d-1}} 
\sqrt{{1\over {f}}  +({\partial_z x^1})^2-f(\partial_z x^0)^2}
\eea  
We shall be  working only for $d> 2$.
From eq.\eqn{schkl1saa} it follows that the 
extremal  surfaces  satisfy following equations
of motion
\bea\label{kl3}
&& {dx^1\over dz}\equiv   {P f\over
 f^{3/2}\sqrt{ {E^2\over f}-P^2+{1\over z^{2d-2}}}} \br
&& {dx^0\over dz}\equiv   {-E \over
 f^{3/2}\sqrt{ {E^2\over f}-P^2+{1\over z^{2d-2}}}} \br
\eea 
The parameters ($E, P$)  are two integration constants. Since we are interested
in knowing the information exchange across the interface,  which is 
  a time dependent phenomenon,
 we will set $x_1 =constant=x_I$
 (along with  
 $P=0$). In this way we are selecting purely time-dependent $(d-2)$ dimensional
extremal surfaces  anchored
at the interface  $x_1= x_I$  at the boundary time $x^0(0)=t_f$. 
The $t_f$ refers to some late time event 
at the boundary. (In  case of the strip
we could  take $ x_1= l/2$ as one edge  of the  strip. 
The same will be true at the other edge also.) 
Then  equations \eqn{kl3}  reduce to a single equation
involving time embedding,
\bea
&& {dx^0\over dz}\equiv   {-E \over
 f^{3/2}\sqrt{ {E^2\over f}+{1\over z^{2d-2}}}} \br
\eea 
Upon integration (and setting $E\equiv 1/z_i^{d-1}$) one obtains
\bea\label{fbn}
 \tau \equiv 
t_f-t_i&\equiv&   \int_{0}^{z_i}  dz {(z/z_i)^{d-1} \over
 f\sqrt{  f + ({z/ z_i})^{2d-2}} } \br
\eea 
where  $t_i$ is an initial time event and given by the 
condition $t_i=x^0(z)|_{z=z_i}$.
 Through above eq.\eqn{fbn} the  constant $z_i$ gets 
related to  time interval $ \tau$ between  two successive
 events occurring at the interface.  
To emphasize, since $|x_0(z)'|\le 1$,  $z=z_i$ is at best
 a {\it cusp}-point where the extremal surface starts
and it ends at $z=0$, see the figure \eqn{fig2} (It is quite unlike in the 
HEE nomenclature involving static RT surfaces, 
there $z_\ast$  rather turns out to be the turning point of the RT surface.). 
Since there is a time translation invariance in the CFT, 
 one may also set  $t_i=0$ and $t_f=\tau$. Thus every where by saying time
we shall only mean the time-interval.

Hence the  net   (entanglement) information exchanged
during  time interval 
$\tau$, across  $x_1=x_I$ interface,
can be obtained by evaluating the area  of extremal 
surface in \eqn{fbn}. From \eqn{schkl1saa}
\be\label{kl1kv}
I_{E}=
{A_{d-2} L^{d-1} \over 4 G_{d+1}}
\int^{z_i}_{\delta}{dz \over  z^{d-1}}{1 
\over \sqrt{f +({z\over z_i})^{2d-2}}} 
\ee  
We can note that the  information exchanged is  dimensionless quantity. 
The $\delta$ acts as cut-off 
to regularize UV divergence in the boundary CFT. 
At $z=z_i$   the time coordinate
on extremal surface   corresponds
to the boundary event $(x^0(z_i)=t_i, x_1=x_I)$. Note $I_E$ is 
proportional to the  
cross-sectional area of the interface between the subsystems 
given by $A_{d-2}=l_2l_3\cdots l_{d-1}$ and it is large.
As the integrand becomes singular as we go near to AdS boundary, 
 we would regularize it by the
contributions of $x^0=constant$, $x^1=constant$ 
  surface and single out the divergent piece.
Thus we obtain
\bea\label{kkl1kv}
I_{E}|_{x_1=x_I}=
&&
{A_{d-2} L^{d-1} \over 4 G_{d+1}}
\left(
\int^{z_i}_{\delta}{dz \over  z^{d-1}}(
{1 
\over \sqrt{f +({z\over z_i})^{2d-2}} } -1) 
-{1\over (d-2) z_i^{d-2}}\right)
+
I_{UV} \br
&&={A_{d-2} L^{d-1} \over 4 G_{d+1}}
{1\over  z_i^{d-2}}\left(
\int^{1}_{0}{d\xi \over  \xi^{d-1}}({1 
\over \sqrt{f +{\xi}^{2d-2}} } -1) 
-{1\over d-2}\right)+ I_{UV}
 \eea  
where  $\xi \equiv {z\over z_i}$, whereas 
the singular UV contribution 
 is  $I_{UV}= {A_{d-2} L^{d-1} \over 4 G_{d+1}}
{1\over (d-2) \delta^{d-2}}$
 which  is positive definite and diverges. It involves only the
area of the interface/boundary. (The UV contribution
 is thus similar to the entanglement entropy in this aspect,
 but it is  independent of local  time information
and forms the universal part.) The first term contains
all time information and it is  finite. 

\subsection{Stop watch approximation}
Let us pick small time interval 
(say the time period of a stop watch). 
We wish to measure net
 information exchanged between two systems staying at the
$x_1=x_I$ interface, and holding a stop watch. 
For  small enough time  the extremal  surfaces \eqn{fbn}
would lie in the vicinity of the 
 asymptotic AdS region only,   we  can easily expect  $z_i \ll z_0$.
In these cases we can evaluate the extremal area  \eqn{kkl1kv} perturbatively
 by expanding it around pure AdS value (i.e. 
treating pure AdS as a ground state of the CFT). 
For small  $\tau$ values, we first   evaluate
 eq.\eqn{fbn} by making  perturbative expansion of the integrand 
\bea \label{dfg1}
\tau
& 
=& z_i \int_0^{1} d\xi
\xi^{d-1}{1 \over  
 \sqrt{R}}\left( 1+ {z_i^d\over z_0^d}(1+{1\over 2R})\xi^{d} 
 +O(({z_i\over z_0})^{2d}) \right)  \br
&&\equiv
z_i \left( b_0 + 
\ep {2b_1 +  i_{11} \over 2}  
+
\cdots\right)
\eea
where   
$\xi \equiv {z\over z_i}$ 
and $R(\xi)\equiv 1+ \xi^{ 2d-2} $.
The dots indicate terms of  higher order  in 
 expansion parameter defined as $\ep=({z_i\over  z_0})^d\ll 1$. 
The  coefficients $b_0, b_1$, and $ i_{11}$  are 
precise integral quantities that  are  
positive definite, but  mostly smaller than unity. These
numbers are  provided in the appendix.
The series \eqn{dfg1} can also be inverted to obtain the
$z_i$-expansion
\bea\label{dfg5}
 z_{i}&=& \bar z_i   
(1 - {\bar z_i^{d}\over  z_0^d} {2b_1+i_{11} \over2  b_0} +\cdots)
\eea
where we define
$\bar z_i= { \t \over  b_0}$ as the value 
specific to  pure AdS  for given   interval $\tau$. 
The equation \eqn{dfg5} summarizes  the net effect of the 
metric deformations (with  black hole)  
on the $z_i$-value perturbatively. One can easily see that
new $z_i$ (with excitations)  is smaller than $\bar z_i$ for  pure AdS. 
Having obtained  $z_i$-expansion,
an expansion  can now be obtained for the  information area functional.
From  the area integral  \eqn{kkl1kv},  
we get the finite part as 
\bea
 I_E 
& =& 
{A_{d-2} L^{d-1} \over 4 G_{d+1}}
 {1\over z_i^{d-2}} \left(\int_0^{1} {d\xi\over \xi^{d-1}}(({1 \over  
 \sqrt{R}}- 1) + {1\over 2}
{z_i^d\over z_0^d}{\xi^{d}\over R^{3\over 2}} 
+\cdots ) -{1\over (d-2)}\right)
\eea
The respective finite integrals can be separately evaluated at 
each order to give 
\bea
 I_E&=&{A_{d-2} L^{d-1} \over 4 G_{d+1}}
{1\over z_i^{d-2}}( -c_0  +  {z_i^{d}\over 2 z_0^d}  c_1   
+\cdots)\br
\eea
where  coefficients 
$c_0, c_1, ... $ are precise real values (see the appendix).
We  emphasize  here that  $c_a$'s are all finite and  positive for  $d>2$.
Substituting $z_i$ expression from \eqn{dfg5} and  keeping terms up to 
first order we  determine 
\bea\label{are1a}
I_E
&&= I_{UV}-{A_{d-2} L^{d-1} \over 4 G_{d+1}}{c_0\over \bar z_i^{d-2}} 
 \left( 1 
+ {\bar z_i^{d}\over  z_0^d}{d-2\over 2} {2b_1+i_{11}\over b_0}-
{\bar z_i^{d}\over 2 z_0^d}  {c_1\over c_0} 
 \right) +\cdots \br
&&\equiv I_{AdS} +I_{(1)}
  \eea
while 
\be
I_{AdS}= 
{A_{d-2} L^{d-1} \over 4 G_{d+1}}
\left({1\over (d-2)\delta^{d-2}} 
-{ c_0b_0^{d-2}\over \t^{d-2}}\right) .
\ee
 is the  information exchanged  in time $ \t$
  for pure $AdS_{d+1}$. It 
 is  positive definite
and is the leading most term
in the  expansion. 
In  large time limit $\t\to \infty$
the $I_{AdS}$ diverges as expected.  
That is to say an infinite information is exchanged by the subsystems
in the CFT ground 
state. In other words
 the {\it ground state constitutes maximum to the information mobility.} 
The subsequent term  in equation \eqn{are1a} 
\bea\label{hhj34ni}
 I_{(1)}&=& 
{A_{d-2} L^{d-1} \over 4 G_{d+1}}
 {c_0\bar z_i^2\over  z_0^d} 
\left(-
(d-2) {2b_1+i_{11}\over 2 b_0}+  {c_1\over2 c_0} 
 \right) \eea
consists  first order contribution of excitations
to the information flow.
The information flow remains maximum for 
pure AdS,
as we  show next, the 
first order contribution (due to excitations) is negative and
tends to reduce the net information flow.  
The 
net change in the information exchange due to excitations is 
\bea\label{hj34ni}
\bigtriangleup I_E  
= - {A_{d-2} L^{d-1} \over 4 G_{d+1}}
{\t^2 D\over  b_0^2  z_0^d} 
\eea
where $D=\left( {-c_1\over2 } + (d-2)c_0 {2b_1+i_{11}\over 2b_0} 
\right)>0$. Note coefficient  $D$ is positive definite  for all $d>2$.
Thus  the expression on the right hand 
side of eq.\eqn{hj34ni} is
negative definite,  suggesting that the net  information 
exchange for the excited CFT has decreased as compared to the
 zero temperature CFT. Also it can be noted that $\bigtriangleup I_E$
 is  quadratic in  time at first order,
 suggesting that having
black hole excitation in the bulk  (or excitations in the 
CFT) decreases net information flow across the subsystem's interface.
 
From eq.\eqn{hj34ni} we may determine the rate  of  reduction in
information flow relative to the vacuum, as
\bea\label{hj4v}
&& 
{\bigtriangleup I_E  
\over \t}  
=- {L^{d-1} A_{d-2}\over 4 G_{d+1}} 
  {\t D \over  b_0^2 z_0^d} 
< 0
\eea
The equation \eqn{hj4v}
represents  net loss per unit time in 
 information mobility due to  excitations. This remains true at least up to first order. 
In summarizing, we have 
learnt that for small time intervals,
\bi
\item The relative {\it loss } in  information exchange 
due to excitations grows quadratically with time. 
\item The information flow is proportional to  cross-section area 
$A_{d-2}$ of the boundary/interface between the systems.
\item It is negative definite for all dimensions indicating that 
 reduction in  information flow  is an universal feature
for excited CFT states. 
\ei
We will show in the next section 
that the higher order corrections, as being perturbatively suppressed,
cannot change this first order leading behaviour of an excited state. 
Nevertheless,
the absolute sign of second order term  will still be  important to know
here.

\subsection{A law of information flow}
The physical quantities such as energy or pressure 
 can be obtained by expanding the bulk AdS geometry \eqn{bst1} in
Fefferman-Graham asymptotic coordinates suitable near the  
boundary \cite{fg},  also in \cite{hs2015}. The
energy density  of the CFT is given by 
\bea\label{hj5}
&& \bigtriangleup {\cal E}=  {  L^{d-1}  \over 16\pi G_{d+1}}  
{d-1\over  z_0^d}\br
\eea
 The {\it pressure}  along all directions is
\bea\label{hj5a}
&& \bigtriangleup {\cal P}=  
  {L^{d-1} \over 16\pi G_{d+1} }  {1 \over z_0^d}
\eea
The  eq. \eqn{hj4v} in terms of physical
 observables may also be written as
\bea \label{alis1}
&& {-b_0^2 \bigtriangleup I_E\over 4\pi  \t}  
=\t A_{d-2} 
  \bigtriangleup {\cal P}_{x_1} 
=
 \t  \bigtriangleup {\cal F}_{x_1} 
\eea
The negative sign indicates that
 there is an overall reduction in information flow or mobility
and it is directly proportional to  the
force (pressure) generated by the excitations
\footnote{
The rate of flow of charge carriers in presence of emf $E$ behaves 
as $j=\sigma E$ in the steady state, where constant $\sigma$ is
the conductivity.}. However
 no  steady state is reached over small times, 
as the rate of information loss  
${ \bigtriangleup \dot I_E}$ only increases with time initially.
However a steady state seems to have been reached after long times, 
where our perturbative approach would rather fail. The numerical plots 
in figures \eqn{fig3a},
\eqn{fig4a},
\eqn{fig5a}
 suggest that late time behaviour generically  
\be
Lim_{\t\to\infty}\bigtriangleup  I_E\to 0
\ee
The above limit may lead to finite answer too. 
Especially in 2-dimensional case the late time behaviour 
can be exactly determined. We
discuss it in the subsection. 

The equation \eqn{alis1}  short of 
 mimics  flow of charge in a conductor under the 
 effect of external emf (force).
In the entanglement context here the eq.\eqn{alis1} thus
 describes   a law of  information flow over small time interval. 
The negative sign is crucial and  universal feature for all $d$. 
We may recall that the entanglement entropy usually 
rises whenever CFT has excitations \cite{hs2015}. This growth in  HEE
 may appear to be related to
 loss in  information exchange between the subsystems. 
Nevertheless it is clear that
 CFT pressure plays vital role in  (entanglement) information exchange 
 (flow).

\subsection{Information at second order and a bound}
In order to make this more robust we need to calculate higher order terms.
Taking steps as in the previous section, we  calculate
the second order terms in the expansion of the information integral, which we
schematically denote as
\be
I\equiv I_{AdS}+I_{(1)} +I_{(2)}+\cdots 
\ee
where $I_{AdS}$ and first order term 
$I_{(1)}$ have already been obtained. We focus on  finding $I_{(2)}$
and its absolute sign in the next. 
In the first step, we obtain  expansion for $z_i$ in terms of $\tau$,  
as done in \eqn{dfg1} and \eqn{dfg5}, up to second order  
\bea\label{dfg52}
z_i&=
&\bar z_i \big[
1 - {(2 b_1 + i_{11})  \over 2 b_0}\bar\ep +
{8 b_1^2 (d+1) + 8 b_1 (d+1) i_{11} + 2  (d+1) i_{11}^2\over 8b_0^2} \bar\ep^2 \br &-& {
   b_0 (8 b_2 + 4 i_{21} + 3 i_{22})\over 8 b_0^2}\bar\ep^2 +O(\bep^3)\big]
\eea
where  $\bar\ep\equiv {\bar z_i^d\over z_0^d}\ll 1$ 
is the expansion parameter. In the next step
the extremal area  calculation leads to the following
second  order term
\newpage
\bea\label{ghy1}
 I_{(2)}&=&{L^{d-1} A_{d-2}\over 4 G_{d+1}} {1\over
\bar z_i^{d-2}}
[{3 c_2 \over 8} +  (d-2) {c_1(2 b_1 + i_{11}) \over 4 b_0} \br
&& - {c_1 d (2 b_1 + i_{11})\over
  4 b_0} - {c_0 \over 8 b_0^2}
  (d-2) (-12 b_1^2 + 8 b_0 b_2 - 4 b_1^2 d - 
12 b_1 i_{11}  \br &&
  -4 b_1 d i_{11} - 3 i_{11}^2 - d i_{11}^2 + 4 b_0 i_{21} + 3 b_0 i_{22})]\bar\ep^2 
\eea
The  parameters $b_a,~i_{ab}$ and $c_a$  have  definite numerical
values provided in the appendix. It is not important to know
them all. A significant result  follows from here is that,   
the $I_{(2)}$ is {\it  positive} definite
 for all $d>2$. This has been thoroughly checked by us. 
The absolute sign of second order term
 is important  as this will provide us with a bound.
It leads to an immediate
conclusion that the loss in  information exchange will have a bound, namely
\bea \label{alis1s}
&& b_0^2{| \bigtriangleup I_E|\over 4\pi \t^2}  
\le   \bigtriangleup {\cal F}_{x_1} 
\eea
where  the net force $ {\cal F}_{x_1} $ is 
given by  cross-section area of the interface times the
entanglement pressure ${\cal P}_{x_1} $ along the $x_1$ direction.
The force is in the transverse direction of the interface
between  subsystems.

To proceed further we now study  individual case of $CFT_d$.
Let us discuss phenomena for $d=3,4,6$ only to gain some 
insight.
 
\ud{Case-I: for $d=4$ $(AdS_5)$}

The  relative information flow per unit area per unit time  
 up to second order is obtained as
\bea\label{hj3c1a}
-{b_0^2\bigtriangleup I_E \over A_2 \t }
=4\pi  \alpha_4 
\t  \left(1-  {.902\over\alpha_4} ({\t \over b_0 z_0})^4
+{2.676\over\alpha_4} ({\t \over b_0 z_0})^8 +O(\bar\ep^3) \right) 
\bigtriangleup P_{x_1} 
\eea
where   numerical values of some coefficients 
for the purpose of estimate are as $\alpha_4\simeq 0.558, ~b_0\simeq .215$. 
The $x_1$ component of  $CFT$ 
pressure exerted on the interface
located at  $x_1= x_I$, is $\bigtriangleup P_{x_1}=
{L^{3} \over 16 \pi G_5} {1\over z_0^4}$ 
 while area of boundary is $A_2$. 
It is  remarkable  that the equation \eqn{hj3c1a} 
is in the factorized form and  may  be written as
\bea\label{hj3c}
-{b_0^2\bigtriangleup I_E
\over A_2 \t}
={4\pi \alpha_4}\t  Q(t,z_0) \bigtriangleup P_{x_1}
\eea
where the factor $Q$  is
\bea
&&
Q=  \left(1  - {.902\over .558} ({\t\over b_0 z_0})^4 +O(\bar\ep^2)
\right) 
\eea
 and it is always  smaller than unity. It may be
  an indication that  effectively the time period 
 (or time per collision) gets  shortened after inclusion
of higher order terms in the perturbative expansion. 
An explicit negative sign on the lhs indicates that there is overall reduction 
in the information flow  for the excited  states. 
As the case here the CFT excitations are thermal, 
but actually that need not have been the case. 
{\it The results would be identical for all finite energy
(IR) gravity perturbations in the bulk.}

However we may define the quantum information 
decay rate per unit  area  as
 \bea\label{hjgc}
-{b_0^2\bigtriangleup I_E\over A_2 \t}
&=&4\pi 
\alpha_4 Q(\t) \t  \bigtriangleup P_{x_1}
 \br & \simeq&  
{4\pi\alpha_4 \over d-1} Q(\t) 
\t  \bigtriangleup {\cal E}
 \br & \le &  
{2\pi\alpha_4 \over d-1}  
\eea
The last inequality follows because $0< Q < 1$,
 and for a quantum state with  energy
 $\bigtriangleup {\cal E}$ and a half-life $\t$
 it is understood that
$\t   \bigtriangleup{\cal E}\simeq {1\over 2}$
due to the  uncertainty.

\ud{Case-2: for $d=3$}
This case involves a 3-dimensional boundary theory which is
 dual of $AdS_4$. 
The  spacetime
$AdS_4$ arises  as near horizon geometry of multiple M2-branes
in $11$-dimensional supergravity. The subsystem interface here is 1-dimensional.
It is found that 
\bea\label{hj3c1}
-{b_0^2 \bigtriangleup I_E \over A_1 \t }
=4\pi  \alpha_3 
 \t  \left(1-  
{.631\over\alpha_3} ({\t \over b_0 z_0})^3
+{1.328\over\alpha_3} ({\t \over b_0 z_0})^6+O(\bar\ep^3) \right)  
\bigtriangleup P_{x_1} \eea
where we have  used  numerical values, 
 $\alpha_3\approx 0.528,~ b_0\simeq .284$. The component of  $CFT_3$ 
pressure $\bigtriangleup P_{x_1}=
{L^{2} \over 16 \pi G_4 z_0^3}$ exerted on the subsystem boundary
located at  $x_1= x_I$, wheras $A_1$ is  length of the  interface.

\ud{Case-3: for $d=6$}
The    information flow per unit area per unit time  
 up to second order for $AdS_7$ bulk spacetime is given by
\bea\label{hj3c1q}
-{b_0^2 \bigtriangleup I_E \over A_4  \t }
=4\pi  \alpha_6 
\t  
 \left(1-  {1.457\over\alpha_6} ({\t \over b_0 z_0})^6
+{6.821\over\alpha_6} ({\t \over b_0 z_0})^{12} +O(\bar\ep^3) \right)  
\bigtriangleup P_{x_1} \eea
where we have  
$\alpha_6\simeq 0.588, ~b_0\simeq .144$. The component of  $CFT_6$ 
pressure $\bigtriangleup P_{x_1}=
{L^5 \over 16 \pi G_7} {1\over z_0^6}$ exerted on the system boundary
located at  $x_1= x_I$, while 4-dimensional subsystem
interface area is $A_4$. 

\section{ Special case of $AdS_3$}

The $AdS_3$ case is  special  where boundary CFT 
is 2-dimensional and bulk geometry is 
the BTZ  black hole. The subsystem interface or junction
is just point-like or a dot. From \eqn{fbn}
we get the slope of the $x^0$ extremal surface
at the cusp point as
\bea 
{t'}|_{z=z_i}= {1\over  (1-{z_i^2\over z_0^2})
\sqrt{2-{z_i^2\over z_0^2}}} \ge {1\over\sqrt{2}}
\eea
that is bounded by its lowest value ${1\over\sqrt{2}}$ for  pure $AdS_3$.
Upon inegration the exact expression for
the cusp value $z_i$ in terms of time gap $\tau$, by integrating \eqn{fbn},
becomes 
\bea
\tanh ({\tau\over z_0})= 
{{z_i\over z_0}(\sqrt{2-{z_i^2\over z_0^2}}-1)\over
1-{z_i^2\over z_0^2}\sqrt{2-{z_i^2\over z_0^2}} } 
\eea
A plot for $\tau~ Vs ~z_i$ has been provided in the figure \eqn{fig21}.
\begin{figure}[h]
\centerline{\epsfxsize=3in
\epsffile{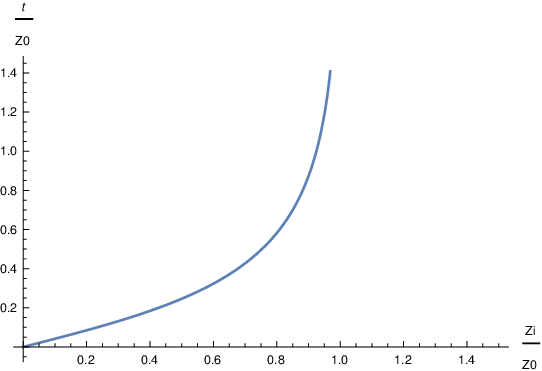} }
\caption{\label{fig21} 
\it The $\tau ~Vs~ z_i$ plot for BTZ case. When $z_i/z_0 \to 1$ the net time required for information processing becomes very large indeed.
We have taken $z_0=1$ here.  }
\end{figure} 
Finally the net information processsed by a quantum dot
 from eq. \eqn{kl1kv} is
\bea\label{3c1n}
  I_E = { L\over 4 G_{3}}\left( 
\ln \left({z_i\over \bar z_i}
{1\over 1+ \sqrt{2-{z_i^2\over z_0^2}}}\right)+ \ln{2\bar z_i\over\delta} 
\right)
\eea
whereas $\bar z_i= { \t \over  b_0}$ is the value corresponding to
pure $AdS_3$
(Note here $b_0=\sqrt{2}-1$). 
For the pure $AdS_3$  (i.e. $CFT_2$ ground state) 
the net information processed in given time interval is given by
\bea
I_{AdS_3}={ L\over 4 G_{3}} \ln ({2\tau \over b_0\delta}).
\eea
These are exact expressions. 

For large  time  ($\tau\to\infty$ )  we  find from \eqn{3c1n}
that
\bea
Lim_{t\to\infty} I_E={ L\over 4 G_{3}} \ln ({z_0 \over \delta})>0 .
\eea
This expression  is finite and completely independent of the time. 
The small variations follow a  relation
 \bea\label{loj}
\delta  I_E \simeq -{ L\over 4 G_{3}} 
{z_0^2\over 2} \delta\left( {1\over z_0^2}\right)
=-{8\over \pi T_{th}^2}\delta {\cal M}  
\eea
where ${\cal M}$ is the energy density of the BTZ 
black hole  and $T_{th}\sim {2\over \pi z_0}$ is horizon
 temperature. {\it It can seen that all local
time dependence  disappears or wiped out after large intervals}.
This large time limit \eqn{loj} may  be 
recasted purely in terms of thermal entropy density as
\bea
 \delta I_E + {8\over \pi T_{th}} \delta s_{th}=0  
\eea
It is an exact relation for the information processed 
by a {\bf quantum dot} (at the interface of the subsystem) and corresponding 
rise in the entropy of any excited state of the system. 
Note here the CFT  is 1-dimensional (spatial) and the system
interface is thus point-like. It suggests that the rise in 
thermal entropy is always
detrimental to the information flow (or processing), read conductivity, 
and vice versa.

\section{Numerical results}

In the previous section we could make perturbative study of the information
flow for finite times only. However the results at large times were elusive because the
 perturbation breaks down. To understand the late time behaviour of information
exchange we pick three cases of $AdS_4$, $AdS_5$ and $AdS_6$. 

For $d=3$ the two plots are obtained in figures \eqn{fig3a} and \eqn{fig3b}. 
\begin{figure}[h]
\centerline{\epsfxsize=3.5in
\epsffile{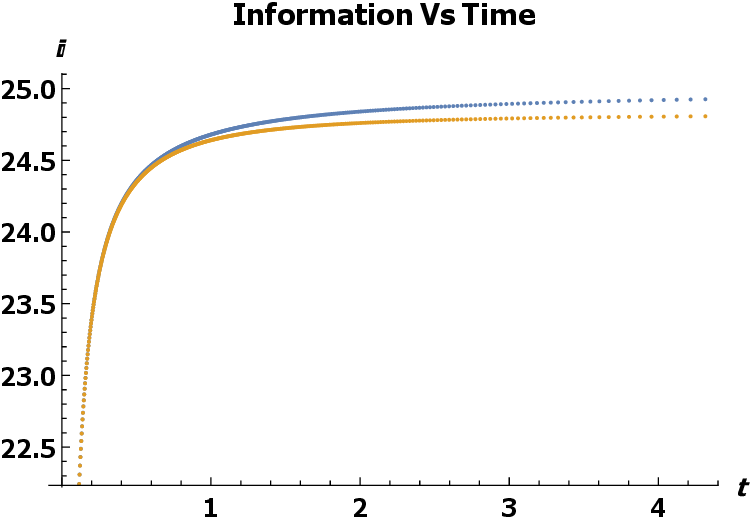} }
\caption{\label{fig3a} 
\it Net information exchange per unit area of the interface 
as a function of time for $d=3$ case. 
We have taken $I_{UV}=25,~z_0=5,~ L=1,~G_N=1/4$. 
The upper curve represents  $I_{AdS_4}$. 
It shows maximum information is exchanged  in the ground state.
 }
\end{figure}
\begin{figure}[h]
\centerline{\epsfxsize=3.5in
\epsffile{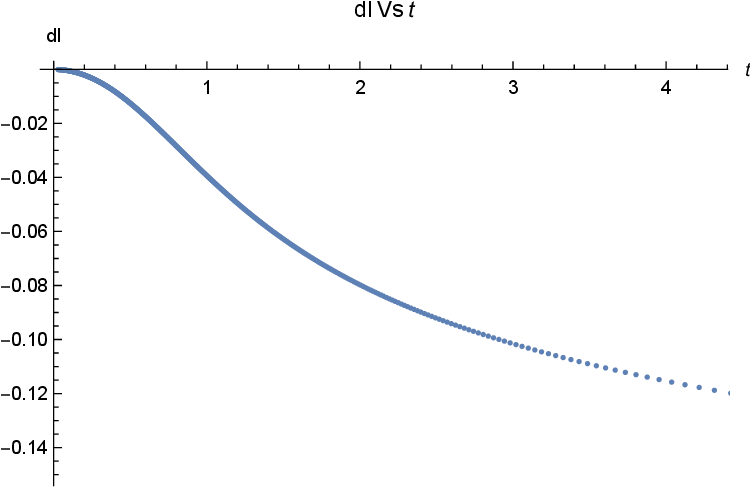} }
\caption{\label{fig3b} 
\it The loss in information flow with excitations for $d=3$. 
We have taken $I_{UV}=25,~z_0=5$. 
  }
\end{figure}

In the fig.\eqn{fig3a} the net information exchange for excited $CFT_3$
rises sharply in tandem with ground state growth and actually 
grows very close up to the ground state plot. But at late time
the two growths differ considerably. The difference of growth 
$\bigtriangleup I$ is plotted in the figure \eqn{fig3b}. 
It depicts a net reduction in the entanglement growth
for the excited state. The information
loss initially grows quadratically in time and then slows down at
late times. The similar results for $CFT_4$ are 
given  in figure \eqn{fig4a} and 
\eqn{fig4b}. 
\begin{figure}[h]
\centerline{\epsfxsize=3.5in
\epsffile{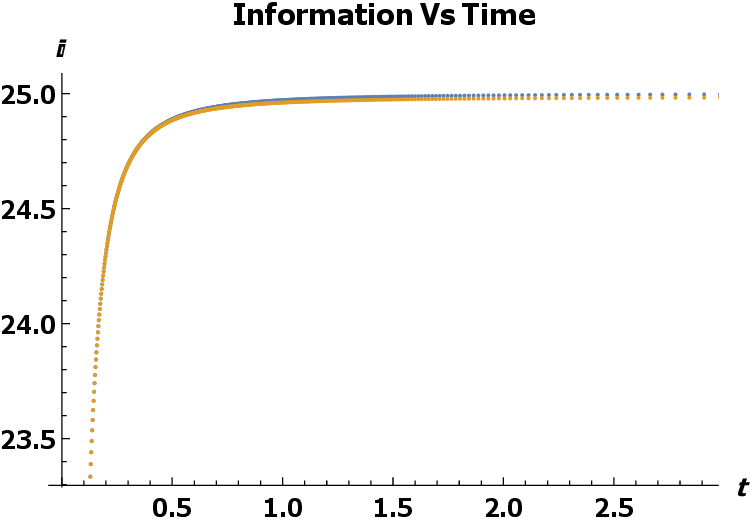} }
\caption{\label{fig4a} 
\it Net information exchange per unit area of the interface 
as a function of time for $d=4$ case. 
We have taken $I_{UV}=25,~z_0=5,~ L=1,~G_N=1/4$. 
The upper curve represents  $I_{AdS_5}$.  
It shows maximum information is exchanged  in the ground state only.
  }
\end{figure}
\begin{figure}[h]
\centerline{\epsfxsize=3.5in
\epsffile{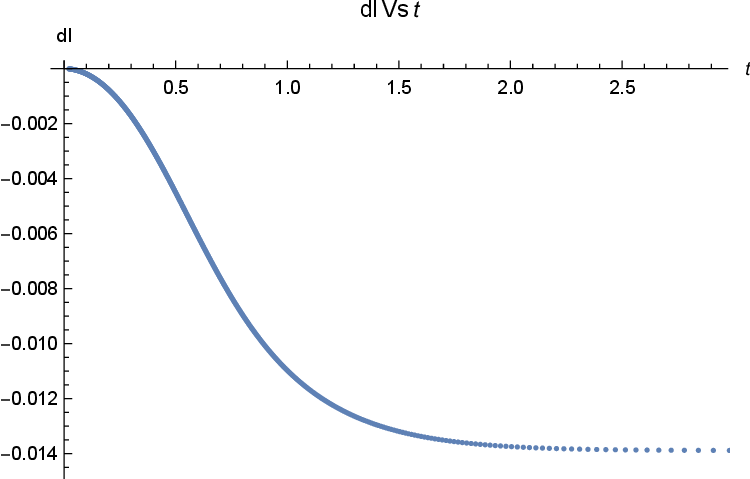} }
\caption{\label{fig4b} 
\it The loss in information flow with excitations for $d=4$. 
We have set $I_{UV}=25,~z_0=5$. 
}
\end{figure}

These graphs show universal feature 
that  information loss (with excitations) grows quadratically near $t=0$,  
but it appears to get saturated at late times and curve gets flattened. 
These results for excited states of $CFT_5$
can be found in figs.\eqn{fig5a} and \eqn{fig5b}. 
There too the information loss curve behaves  quadratically
 in time initially but unlike previous two cases it rebounds 
and information loss
starts reducing at late times. This is a direct effect of dimensionality
of the theory.  The early time behaviour is consistent 
with  perturbative analysis where first order term 
is indeed quadratic in time.   
\begin{figure}[h]
\centerline{\epsfxsize=3.5in
\epsffile{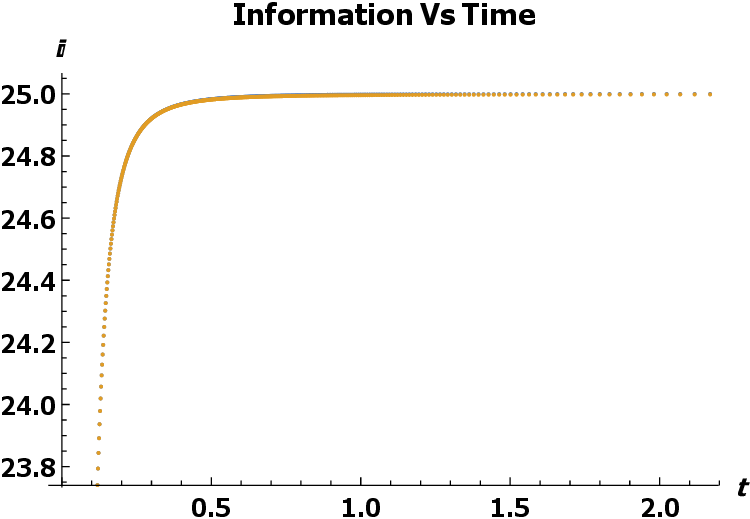} }
\caption{\label{fig5a} 
\it Net information exchange 
per unit area of the interface 
as a function of time for $d=5$ case. 
We have taken $I_{UV}=25,~z_0=5,~ L=1,~G_N=1/4 $. 
The upper curve represents  $I_{AdS_6}$.  
It shows maximum information is exchanged  in the ground state.
  }
\end{figure}
\begin{figure}[h]
\centerline{\epsfxsize=3.5in
\epsffile{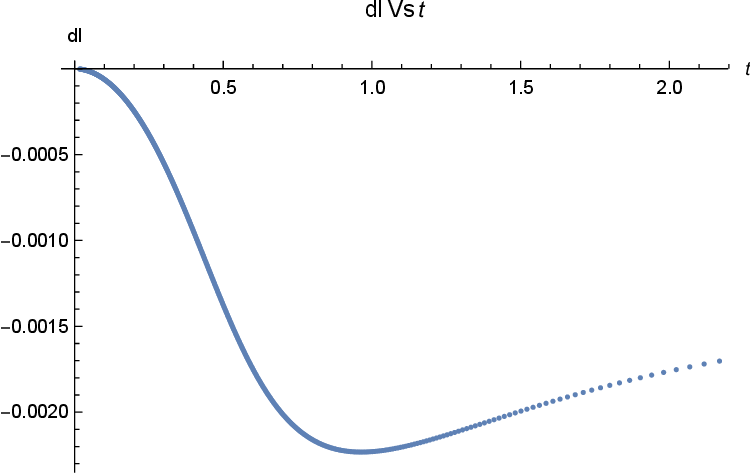} }
\caption{\label{fig5b} 
\it The loss in information flow with excitations for $d=5$. 
We have taken $I_{UV}=25,~z_0=5$.   }
\end{figure}

\section{Quantum  Complexity}
We now evaluate another time dependent quantity when the
CFT is in excited state. The 
complexity of quantum circuits is a 
measure of time evolution of the state of the 
system as a whole. It can  be defined 
holographically  by the volume action functional describing a time-dependent
 (codim-1) extremal surface \cite{susski}
\bea\label{schkl1sac}
 C_E  &\equiv& 
 { V_{d-1}  L^{d}\over 4 G_{d+1}}
\int {dz\over z^{d}}  
\sqrt{{f}^{-1}  -f (\partial_z x^0)^2}
\eea  
where  
 $V_{d-1} \equiv  l_1 l_2 l_3 \cdots l_{d-1}$ is 
the net spatial volume of the CFT. In this sense quantum
complexity is a bulk property of the CFT and extensive  in nature. 
Note the difference that the volume complexity is  dimensionful  
whereas information exchange $I_E$  is dimensionless.
From \eqn{schkl1sac} it follows that  
 extremal  surfaces have to satisfy following equation of motion 
\bea\label{kl3c}
&& {dx^0\over dz}\equiv   { z^d/
z_c^d\over f\sqrt{ f+  z^{2d}/z_c^{2d}}} \br
\eea 
The  $z_c$ is the integration constant and it 
will get related to the time interval on the boundary. 
To first solve  $x^0$ equation, 
let us choose an extremal  surface having
  boundary value $x^0(0)=t_f$, where $t_f$ is a time event  
on the boundary. The eq.\eqn{kl3c}
 gives upon integrating
\bea\label{lo9}
 \t =t_f-t_i&\equiv&   \int_{0}^{z_c}  dz {(z/z_c)^{d} \over
 f\sqrt{  f + ({z/ z_c})^{2d}} } \br
\eea 
where $t_i$ is some initial time  given by $t_i=x^0|_{z=z_c}$.
 Thus the  constant $z_c$ gets 
related to time interval between two events on the boundary. 
There is  overall time translation
invariance in the system. It is also clear that for  small
 intervals we will have a situation where $z_c \ll z_0$. 

The  expression for    complexity 
during time  interval 
$\t$, can be obtained by calculating the area  of the  
extremal surface from action \eqn{schkl1sac}
\be\label{kl1kvc} 
C_{E}=
{V_{d-1} L^{d} \over 4 G_{d+1}}
\int^{z_c}_{\delta}{dz \over  z^{d}}{1 
\over \sqrt{f +({z\over z_c})^{2d}}} 
\ee  
The integrand in \eqn{kl1kvc} becomes singular  
near the AdS boundary which is 
the UV divergence of the CFT. 
We can regularize it by the contribution of $x^0=constant$
surface and single out the diverging UV part.
Thus we obtain
\bea\label{kl2kvc}
C_{E}=
{V_{d-1} L^{d} \over 4 G_{d+1}}
{1\over  z_c^{d-1}}\left(\int^{1}_{0}{d\xi \over  \xi^{d}}({1 
\over \sqrt{f +{\xi}^{2d}} } -1) 
-{1\over d-1}\right)+ C_{UV}
 \eea  
The  divergent
part is given by $C_{UV}= {V_{d-1} L^{d} \over 4 G_{d+1}}
{1\over (d-1) \delta^{d-1}}$,
  it is positive definite and  proportional to  spatial
volume of the CFT. We will now evaluate $C_E$ perturbatively for small $\tau$.  
First, by evaluating rhs of eq.\eqn{lo9} up to first order, 
we find  
\bea\label{ghb}
z_c=\bar z_c(1- 
 {2f_1+j_1\over2 f_0}\bar\ep  + O(\ep^2))
\eea
where 
$\bar z_c= {\t\over f_0}$ and expansion parameter
$\bar\ep={\bar z_c^d\over z_0^d}\ll 1$.
The coefficients $f_0,~ f_1$ and $j_1$  are all finite and
 given in the appendix.
It is not immediately important to know them here. 
Using $z_c$ equation \eqn{ghb} and substituting in \eqn{kl2kvc}, we find
the net change in complexity $ (C_E-C_{AdS})$ to be given as
 \bea\label{hj34}
\bigtriangleup C_E
&=& 
{V_{d-1} L^{d} \over 4 G_{d+1}}
 {d_0 \bar z_c\over  z_0^d} 
\left( -(d-1) {2f_1+j_1\over 2 f_0}+  {d_1\over2 d_0}  \right) + higher~ orders
\br
&&= -{V_{d-1} L^{d} \over 4 G_{d+1}}
{\t \over  f_0}
{D_c\over  z_0^d} + higher~ orders
\eea
Thus  up to first order 
the change in complexity per unit volume
per unit time  is
 \bea\label{hj34v}
{f_0 \bigtriangleup C_E\over V_{d-1}\tau}
&\simeq& -{4\pi D_c }
   \bigtriangleup {\cal{E}}
\eea
where 
$$D_c\equiv  (d-1) d_0{2f_1+j_1\over 2 f_0}-  {d_1\over2 }   $$
 and  note that $D_c>0$
 for all $d$ values. For example, for $d=4$ one gets $D_c\approx .432$
and $f_0\simeq .173$. Thus
the  negative sign in \eqn{hj34v}  indicates that overall  
holographic complexity (HC) of the excited CFT  will get reduced 
 as compared to the ground state (i.e. pure AdS). 
This reduction in circuit complexity  is analogous 
 to the reduction in information exchange 
 at the interface  in previous section. 
While, in contrast, we may recall here that entanglement entropy 
(HEE)  increases when the CFT  is excited.
We  interpret the two opposing natures of  HC and HEE, 
by suggesting that complexity is a measure of (global) 
symmetries of a given quantum state. The complexity will get reduced with  
 increased symmetry of a quantum state. But
 excitations are also cause of rise in the (internal) disorder in the system, 
 whence the HEE will essentially increase. But amidst disorder there 
can still be an overall enhancement of the global symmetry in the system 
(such as rotations), and therefore the HC would decrease. 
It is very much like the phenomena we 
 encounter in ferro-magnetic (spin) systems in three dimensions.

\subsection{A  bound on loss of information exchange}
The bound in the last section on relative information flow can also 
be stated  differently in terms of complexity as (in $d>2$) 
\bea \label{alis1s2h}
&& 
{ b_0^2 \bigtriangleup \dot I_E\over A_{d-2} }  
\le {2 \alpha_{d}\over L D_c}
{ f_0\bigtriangleup C_E\over V_{d-1} }  
\eea
where $A_{d-2}$ is the area of the system's interface and 
$V_{d-1}$ is total volume of the system. $\bigtriangleup C_E$
measures the net loss in quantum complexity. The  constants
$\alpha_d$ and $D_c$ depend on  spacetime dimensions.
(Especially for $d=4$, we  have  
$\alpha_{4}\simeq .558$ and $D_c\simeq.432$.) 
The inequality \eqn{alis1s2h} can be obtained for any excited  CFT state. 
It states that rate of loss in information flow due to the excitations 
remains bounded
by the corresponding loss in the quantum complexity of the given state.
However, as we notice both quantities suffer loss in the presence of 
the excitations. The excitations in the present case are thermal due to
the black holes in the bulk. But the above relation
 would remain true for any positive energy
fluctuations also.

\section{Summary}
We explored net quantum
  information exchange or sharing  in real time  at the interface 
between  subsystems for  $CFT_d$ living at the boundary of 
$AdS_{d+1}$ spacetimes. The information exchange
is a continuous phenomenon and only after long time intervals
full information exchange gets saturated. Generally we prepare systems
for small time intervals only. We find perturbatively that the rate of 
information exchange  between two systems 
is always reduced when the systems have excitations for 
all $d$, while 
CFT ground state has a maximum information exchange rate. 
 The net information loss  is directly proportional to  the
force  or pressure generated by the CFT excitations.
We get a bound
$$
 b_0^2{| \bigtriangleup I_E|\over 4\pi \t^2}  
\le   \bigtriangleup {\cal F}_{x_1} 
$$
where  the  force $ {\cal F}_{x_1} $ is in transverse 
direction to the interface. 
The inequality is saturated only at the first order of perturbation.
Especially for 2-dimensional CFT, the large time relation  
can be recasted in terms of thermal entropy density as
$$
  \delta I_E + {8\over \pi T_{th}} \delta s_{th}=0  
$$
It is an exact relation for the loss in information processed 
by the {\bf quantum dot} (the interface) and corresponding 
rise in thermal entropy for the excited state of the system. 
Note  the $CFT_2$ system represents 1-dimensional wire 
and the subsystem interface is only  point-like. 
It proves that the rise in entropy is always
detrimental to the information flow (or conductivity) and vice versa.

We further observed that there is an 
overall reduction in  quantum complexity too in the presence of excitations.  
While it is well understood that
excited states add more to the entanglement entropy due to rise in disorder. 
From this perspective
we are led to argue that the reduction in the  complexity may be
due to relative higher symmetry excited states may have 
as compared to  CFT ground state. 
The ground state is always more ordered but it maybe less symmetric 
as we observe in ferromagnetic systems with an spontaneously broken
symmetry. 

We have got a new bound which
states that rate of loss in the information flow due to excitations 
will remain bounded
by the corresponding loss in quantum complexity for a given state
of the system.

Before we conclude there are a few clarifications to add. Most of our
computations are on the gravity side assuming standard holographic method.
It would be challenging to develop same perspective on the boundary side. We
have observed  that information exchange is generally 
reduced in the presence of excitations in the
CFT as compared to the ground 
state (at zero temperature). This is found to be true in all dimensions. 
The reduction in information exchange
 is also one of the common characterisitic in most 
 systems (classical or quantum) found in nature. 
As an example the transport in metals (a type of information flow) 
 is also reduced at high temperatures and 
in the presence of impurities too. It would be interesting
to understand the information processing of quantum computers
in real time which work on the principle of quantum teleporation.
The quantum-dot in section-3 is an interface or junction of 1-dimensional wire like system and it works quite like a quantum computer. It processes the information
lying on two sides of the wire.
Furthermore, we have only considered black holes as excitations
on top of pure AdS geometry. This was for simplicity purposes only.
However the calculations are mostly perturbative 
and results depend upon  dimensionless ratio, $z_i/z_0$, 
which is very small than unity. So we are studying boundary phenomena
over small time intervals  for which the extremal surfaces 
reside in the asymptotic region only. 
The actual black hole (IR) geometry does not 
play significant role in it. This indicates that other small excitations of the AdS 
geometry can also be studied and we believe the outcome would be similar.

 

\vskip.5cm
   
\appendix{

\section{Integral Quantities}

Some useful  integral values we have used in the calculation of information 
flow are being noted down here
(where $R(\xi)=1+\xi^{2d-2}$) 
\bea
&& b_0=\int_0^{1} d\xi \xi^{d-1}{1 \over   \sqrt{R}}
\simeq .284 ~({\rm for}~d=3) 
\br
&& b_1=\int_0^{1} d\xi \xi^{2d-1}{1 \over   \sqrt{R}}
\simeq .133~ ({\rm for}~d=3) \br
&& b_2=\int_0^{1} d\xi \xi^{3d-1}{1 \over   \sqrt{R}}
\simeq .086 ~({\rm for}~d=3) \br
&& i_{11}=\int_0^{1} d\xi \xi^{2d-1}{1 \over R  \sqrt{R}}
\simeq .087 ~({\rm for}~d=3) \br
&& i_{21}=\int_0^{1} d\xi \xi^{3d-1}{1 \over R  \sqrt{R}}
\simeq .052 ~({\rm for}~d=3) \br
&& i_{22}=\int_0^{1} d\xi \xi^{3d-1}{1 \over R^2  \sqrt{R}}
\simeq .033 ~({\rm for}~d=3) 
 \br
&& c_0=\int_0^{1} d\xi \xi^{1-d}(1- {1 \over   \sqrt{R}}) +{1\over d-2}
\simeq 
1.131 ~({\rm for}~d=3) 
\br
&& c_1=\int_0^{1} d\xi {\xi \over R  \sqrt{R}}
\simeq .353 ~({\rm for} ~d=3) \br
&& c_2=\int_0^{1} d\xi {\xi^{d+1} \over R^2  \sqrt{R}}
\simeq .077 ~({\rm for}~ d=3), 
\eea
We have provided numeric
values for $d=3$ case
only. For other $d$ cases one can obtain these values easily.

Another set of  integral coefficients which appear in complexity are 
(with $R(\xi)=1+\xi^{2d}$) 
\bea
&& f_0=\int_0^{1} d\xi \xi^{d}{1 \over   \sqrt{R}}
\simeq .214 ~({\rm for}~d=3) 
\br
&& f_1=\int_0^{1} d\xi \xi^{2d}{1 \over   \sqrt{R}}
\simeq .117 ~({\rm for}~d=3) 
 \br
&& j_1=\int_0^{1} d\xi \xi^{2d}{1 \over R  \sqrt{R}}
\simeq .080 ~({\rm for}~d=3) 
 \br
&& d_0=\int_0^{1} d\xi \xi^{-d}(1- {1 \over   \sqrt{R}}) +{1\over d-1}
\simeq 
0.599 ~({\rm for}~d=3) 
\br
&& d_1=\int_0^{1} d\xi {1 \over R  \sqrt{R}}
\simeq 0.867 ~({\rm for} ~d=3) \br
\eea
while
\bea
&& f_0=\int_0^{1} d\xi \xi^{d}{1 \over   \sqrt{R}}
\simeq .173 ~({\rm for}~d=4) 
\br
&& f_1=\int_0^{1} d\xi \xi^{2d}{1 \over   \sqrt{R}}
\simeq .091 ~({\rm for}~d=4) 
 \br
&& j_1=\int_0^{1} d\xi \xi^{2d}{1 \over R  \sqrt{R}}
\simeq .063 ~({\rm for}~d=4) 
 \br
&& d_0=\int_0^{1} d\xi \xi^{-d}(1- {1 \over   \sqrt{R}}) +{1\over d-1}
\simeq 
0.413 ~({\rm for}~d=4) 
\br
&& d_1=\int_0^{1} d\xi {1 \over R  \sqrt{R}}
\simeq 0.107 ~({\rm for} ~d=4) \br
\eea
We have only provided these numerical values 
just for having an idea.

}


\end{document}